\documentclass[final,1p,times]{elsarticle}
\usepackage{amssymb}
\usepackage{amsmath}

\journal{Photoacoustics}

\begin{document}

\begin{frontmatter}

\title{Demonstration of continuous non-invasive monitoring of plant metabolism using mid-infrared photoacoustic spectroscopy\tnoteref{t1}}                      
\tnotetext[t1]{This work is part of the IPCEI Microelectronics and Connectivity and was supported by the French Public Authorities within the frame of France 2030, the CARNOT Institute LETI and the French National Research Agency (ANR) through the project the EXPLOR'AE program HARMONICA (ANR-24-RRII-0003)}

\author[1]{Ellyne Liégeois}
\ead{ellyne.liegeois@cea.fr}
\cortext[cor1]{Corresponding author}

\author[1]{Héloïse Hannart}

\author[1]{Jules Skubich}

\author[1]{Maeva Doron}

\author[1]{Kevin Jourde}

\author[2]{Johan Decelle}

\author[3]{Yann Bresson}

\author[3]{Sylvain Delzon}

\author[3]{Régis Burlett}

\affiliation[1]{
organization={Univ. Grenoble Alpes, CEA, LETI},
addressline={F38054}, 
city={Grenoble},
country={France}
}

\affiliation[2]{
organization={Cell and Plant Physiology Laboratory, CNRS, CEA, INRAE, IRIG, Univ. Grenoble Alpes},
addressline={F38054}, 
city={Grenoble},
country={France}
}

\affiliation[3]{
organization={Univ. Bordeaux, UMR 1202 BIOGECO, },
addressline={F33400}, 
city={Bordeaux},
country={France}
}

\begin{abstract}

Life is a rhythmic process driven by cellular and physiological mechanisms that allow plants to respond and adapt to their environment. In addition to natural fluctuations, plants are increasingly exposed to environmental stresses, mainly caused by climate change and anthropogenic pollution. Multiple approaches across different biological scales (gene, cell, metabolite, and physiology) have shed light on acclimation mechanisms but generally lack sufficient temporal resolution. Establishing a non-invasive and high-temporal-resolution monitoring method to measure plant activity across different conditions is therefore needed to fully assess how plant metabolism changes under stress. Here, we established a novel approach to perform continuous and non-invasive mid-infrared photoacoustic spectroscopy on vascular plants, with the aim of identifying and monitoring the chemical composition of peripheral tissues, including sap, over periods ranging from a few minutes to several months. We describe an experimental setup, the physics underlying its operation, and present a multi-day acquisition. Compared with controls, data enabled to identify a diurnal pattern, with peaks of approximately equal duration, that recurred consistently over the experiment. Estimations of the impact of instrumental and environmental fluctuations indicate that the signal amplitude and trends cannot be explained by instrumental drifts. The experiment thus demonstrated the detection of a signal associated with plant physiological activity. While its origins remain to be clarified, this preliminary experiment demonstrates the feasibility of the measurements and provides a basis for future developments.

\end{abstract}

\begin{graphicalabstract}
\includegraphics[width=1.00\linewidth]{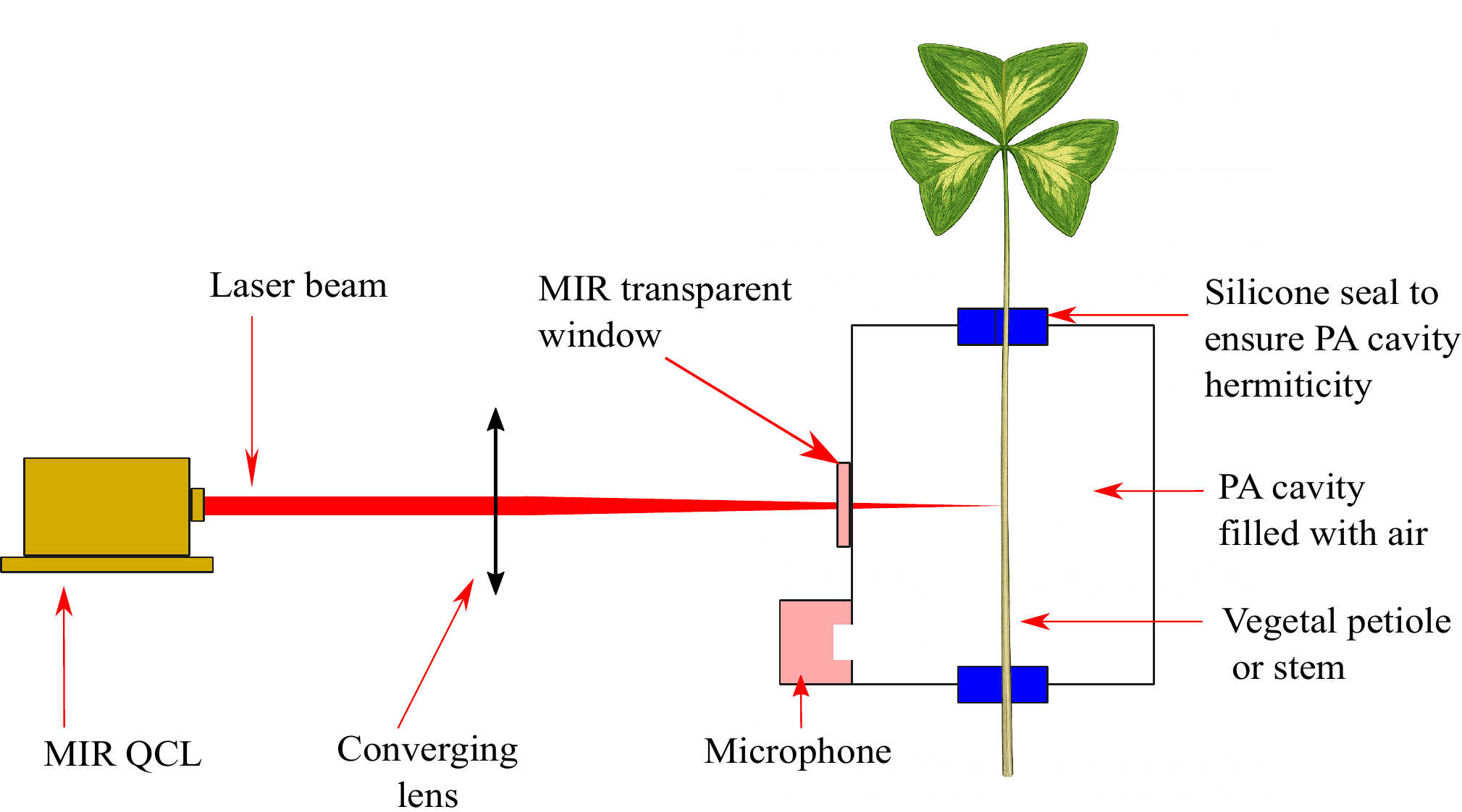}
\end{graphicalabstract}

\begin{highlights}
\item Description of an infrared spectroscopy photoacoustic cavity adapted to plant stems.
\item Execution of a continuous, multi-day experiment on a living plant sample.
\item Detection of a signal induced by plant physiological activity.
\end{highlights}


\end{frontmatter}

\section{Introduction}
\label{sec:introduction}

Plants growth, development, and responses to environmental changes strongly rely on the production, transport, and use of metabolites (carbohydrates, amino acids, organic acids, lipids and phytohormones \cite{lazar_taiz_2003}). Together, these molecules sustain virtually every aspect of plant physiology: from photosynthesis and respiration to growth, reproduction, flowering and stress acclimation. Their production and fluxes are driven by daily rhythms that vary with the seasons \cite{farre_interactions_2012, kim_importance_2017}. Alterations of varying intensity related to environmental disturbances then superimpose on biological cycles. In the era of human-induced climate change, environmental stresses are becoming more intense, frequent and unpredictable \cite{intergovernmental_panel_on_climate_change_ipcc_climate_2023}. Consequently, the ability to monitor plant metabolic dynamics at relevant spatial and temporal resolutions \textit{in vivo} is essential to improve our understanding of metabolic fluxes in plants as well as their adaptive response to environmental stresses. 

\begin{figure}[htbp]
\centering
\includegraphics[width=\linewidth]{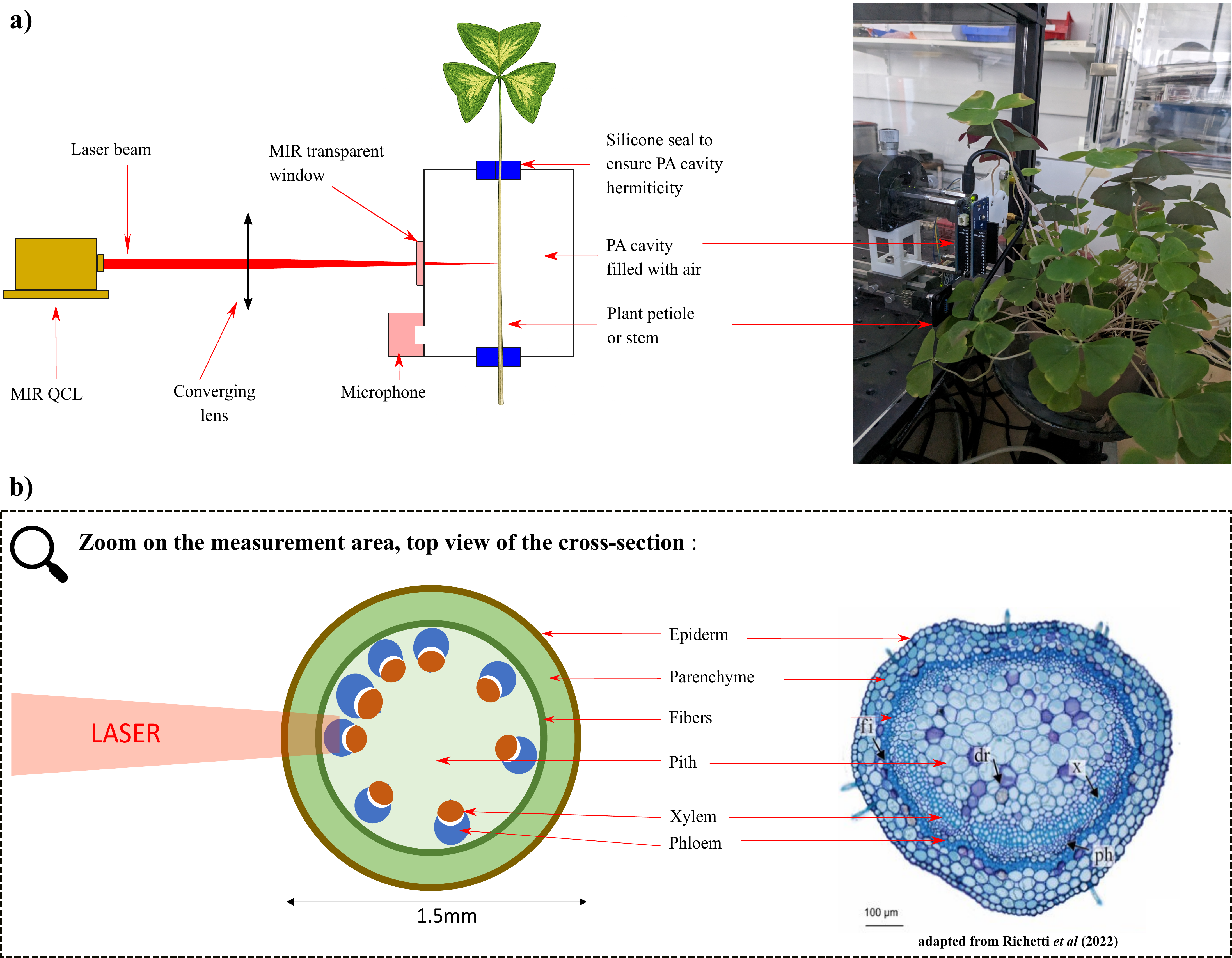}
\caption{Subfigure \textit{a}~: experimental setup scheme and picture. An IR beam is generated by a QCL and focused on the plant stem or petiole. The sample is imprisoned inside an air cavity, which is closed by optimized seals matching the stem geometry, a MIR transparent window to transmit the IR beam to the stem, and an electronic card equiped with a microphone. The air cavity converts temperature variations induced by photothermal effect into pressure variations, measured by the microphone. Subfigure \textit{b}~: A close up on the tested plant, an \textit{Oxalis Triangularis} petiole, is depicted in section view. The laser optical depth penetration is suitable to reach the phloem. The \textit{Oxalis Triangularis} cross section picture is adapted from \cite{richetti_macro-_2023} with the authors agreement.}
\label{fig:oxalis_and_setup}
\end{figure}

Phloem is the principal long-distance transport network of vascular plants, distributing elaborated sap from photosynthetic leaves to roots \cite{schulz_phloem_2009}. This sap contains carbon energy (e.g. glucose and various other carbohydrates)  \cite{dinant_phloem_2010, hijaz_collection_2014}, and numerous molecules including proteins and RNA \cite{mccubbin_phloem_2021}. Phloem tissue is located under the bark at the periphery of stems, $5~\mathrm{\mu m}$ to $3~\mathrm{mm}$ below the surface. Depending on the plant species, it may appear as a continuous layer exhibiting radial symmetry around the stem, or as bundles distributed around the periphery of the stem \cite{hardtke_phloem_2023}. Its thickness ranges from a few tens of micrometers to several millimeters depending on the plant species and age (see Fig.~\ref{fig:oxalis_and_setup} for the \textit{Oxalis Triangularis} example, with vascular bundles). Sap transport occurs according to the Münch pressure-flow mechanism through a network of highly specialized sieve elements \cite{munch_stoffbewegungen_1930, de_schepper_phloem_2013}, whose organization is remarkably conserved across vascular plants \cite{jensen_universality_2012}. The function of phloem thus make this tissue a preferential target for studying the dynamics of primary metabolites. 

To study phloem metabolites, nearly all standard analytical techniques rely on destructive tissue sampling followed by laboratory analyses (liquid or gas chromatography often coupled with mass spectrometry, Raman/Infrared(IR) spectroscopy, enzymatic assays, etc). In addition to requiring a heavy workload, these approaches exhibit many limits. They provide temporally discrete measurements and require tissue destruction, preventing monitoring over time by repeated measurements on the same organ or tissue. Non-destructive methods also exist. CEST- MRI (chemical exchange saturation transfer - magnetic resonance imaging) enabled the monitoring of sugars and amino acids in seeds over sixty hours \cite{mayer_metabolic_2024} and within a sugarcane stem. While interesting, this solution is expensive and difficult to deploy in outdoor environments. A few recent methods based on electrochemical reactions enabled continuous monitoring of a few metabolites \textit{in vivo} in plants \cite{altangerel_vivo_2017, zhang_pilot_2023}, but they remain invasive. Developing non-invasive long-term approaches capable of continuously monitoring metabolites in plant phloem, and with a potential for outdoor measurements, therefore represents a major technological challenge.

Optical vibrational spectroscopy has proven to be a relevant method for non-invasive chemical sensing \cite{juarez_contemporary_2024} thanks to its remarkable molecular specificity for a wide range of biological compounds. Mid-infrared (MIR) spectroscopy specifically can rely on the continuous improvement of high-quality MIR optical sources (Quantum Cascade Lasers (QCL) \cite{coutard_volume_2020, lepage_hybrid_2025}) unlocking sensitivity gains and encouraging sensor miniaturization. The standardization of MIR spectroscopy components has been driven over the last decade by the quest for non-invasive monitoring of various human physiological parameters, especially the glucose level \cite{kottmann_mid-infrared_2016,golparvar_rapid_2022, coutard_neogly_2024,uluc_non-invasive_2024}. However, to our knowledge, these technologies have not yet been adapted for long-term plant monitoring. Achieving this objective requires overcoming several plant-specific challenges such as long-term mechanical stability, complex signal processing and data interpretation of heterogeneous biological tissues.

To tackle this challenge, we designed a mid-infrared photoacoustic spectroscopy setup adapted to plant stems and petioles. With our new design, a multi-day monitoring experiment demonstrates the ability to record biologically related signals with specific temporal patterns. This proof-of-concept setup, which needs to be complemented by additional analyses to enable unambiguous biological interpretation, paves the way for future developments to properly assess plant physiology in natural and stress conditions over time.

\section{Measurement principle and notations}

\subsection{Photoacoustic Infrared Spectroscopcy : measurement principle}
\label{subsec:pa_ir_s}

The plant monitoring method introduced in this paper is based on indirect photoacoustic MIR spectroscopy (IPAS), described below.

\subsubsection*{Infrared Spectroscopy}

IR spectroscopy is a widely used laboratory technique for characterizing the molecular composition of matter in the gas, liquid, and solid states. \cite{griffiths_introduction_2007}. It relies on molecules specific vibrational response to light in the $5-25~\mathrm{\mu m}$ electromagnetic spectral range. The composition of unknown samples may be quantitatively identified using large spectral databases either simulated or acquired from reference samples. There are multiple techniques to perform IR spectroscopy, and a method can be selected depending on the experimental constraints. IR spectroscopy requires an IR optical source to probe the sample, and a transduction technique to measure how the light is absorbed. The most common laboratory technique is Fourier Transform IR (FT-IR) spectroscopy \cite{baker_using_2014}, relying on a black-body source, and a Michelson interferometer with an adjustable optical path length to characterize the light either transmitted through, reflected by, or scattered by the sample.

\subsubsection*{Photothermal Spectroscopy}

There are also indirect transduction techniques, in the sense that they do not characterize the light that interacted with the sample, but how the sample reacts to light absorption. Photothermal spectroscopy is one of these techniques. It consists in detecting the heat released by the probed sample as a proxy of its optical response \cite{rosencwaig_theory_1976, mcdonald_generalized_1978}. There are multiple ways of probing heat \cite{tam_applications_1986}, among which the most popular are direct photoacoustic (PA) spectroscopy (measurement of the acoustic wave generated by the sample mechanical dilatation), deflection spectroscopy (measurement of the optical deflection of a second optical beam in a secondary material set on the sample, and whose refractive index is sensitive to temperature changes), and indirect PA spectroscopy (IPAS). IPAS consists in using a secondary material, usually a gas enclosed in a cavity, which is set on the sample and converts temperature variations triggered by heat release in pressure variations. This is the approach adopted in this manuscript.

\subsubsection*{Advantages and Drawbacks of Photothermal Spectroscopy}
\label{subsec:advantages_and_drawbacks}
Photothermal transduction techniques distinguish themselves by numerous features. The first is their ability to probe optically opaque or diffusive samples, as there is no need to recover the incident light after its interaction. Besides, photothermal techniques are sensitive to the shape of the heat source, which follows the Beer-Lambert law in a non-diffusive sample. This law states that in a homogeneous sample characterized by its absorption coefficient $\beta~\mathrm{[m^{-1}]}$, the laser intensity $I~\mathrm{[W.m^{-2}]}$ reads:

\begin{equation}
I (\lambda) = I_0 \mathrm{e}^{-\beta(\lambda) z} ~~~\mathrm{,}
\label{eq:beer_lambert}
\end{equation}
where $\lambda~\mathrm{[\mu m]}$ is the optical source wavelength, $z~\mathrm{[m]}$ the distance from the sample surface, and $I_0 = I(z = 0)$. For water in the $7-10~\mathrm{\mu m}$ spectral range, $\beta \sim 500~\mathrm{cm^{-1}}$. This means that $95~\%$ of the heat source is contained within $\sim 3 / \beta = 60~\mathrm{\mu m}$. In other words, the photothermal signal is mostly sensitive to the first $60$ micrometers of the sample. For a biological sample such as a tree stem, Eq.~\ref{eq:beer_lambert} is not directly valid as the sample is heterogeneous. The first layer of a tree, the bark, should also be drier than underneath living tissues. As a consequence, it must be possible to probe the stem deeper, typically down to $100$ to $200~\mathrm{\mu m}$. This feature is advantageous, for example, compared with attenuated total reflection (ATR) spectroscopy, which is limited to characterizing depths on the order of the optical source wavelength. 

Another characteristic feature of most photothermal transduction schemes is the possibility to obtain additional information about the sample by tuning the modulation frequency of the optical source intensity $f~\mathrm{[Hz]}$. Indeed, the PA cavity converts into pressure the heat that manages to diffuse from the photothermal source to the sample surface. This characteristic distance is the sample thermal diffusion length $\mu~\mathrm{[m]}$ defined as: 

\begin{equation}
\mu = \sqrt{\frac{\alpha}{\pi f}} ~~~\mathrm{,}
\label{eq:mu_def}
\end{equation}
where $\alpha~\mathrm{[m^2.s^{-1}]}$ is the sample diffusivity. In pure water, $\mu \sim 220 / \sqrt{f ~ \mathrm{[Hz]}} ~\mathrm{\mu m}$. There is a characteristic frequency $f_t$ at which the optical penetration length $l_{opt}~\mathrm{[m]} = 1 / \beta$ is equal to the thermal diffusion length:

\begin{equation}
f_t = \frac{\alpha}{2 l_{opt}^2} ~~~\mathrm{.}
\label{eq:ft_def}
\end{equation}
If $f \ll f_t$, the thermal diffusion length is very large with respect to the optical absorption length $l_{opt}$, making the photothermal measurement independent from the sample optical properties. The signal thus only depends on the sample thermomechanical properties. If $f \gtrsim f_t$, only part of the photothermal heat is recovered, and the higher $f$ is, the smaller the region from the heat source that contributes to the signal. The signal depends both on the sample thermomechanical and optical (intrinsically linked to chemical) properties. As a consequence, by changing $f$, it is possible to tune the measurement sensitivity to the sample optical response, as well as to perform a rudimentary depth scan at the optical penetration length scale. In pure water, $f_t \sim 120 ~\mathrm{Hz}$, meaning that for acquisition on biological samples, $f$ typically needs to be in the range of $10~\mathrm{Hz}$ to $1~\mathrm{kHz}$. On the one hand, this complexity means the measurement provides a rich information, but on the other hand, it makes interpretation complex and often necessitates the use of models and hypotheses about the sample structure and properties to extract the information of interest. 

In addition, IPAS is well suited to performing photothermal spectroscopy on uneven or geometrically unstable surfaces. Indeed, the air in the PA cavity, which constitutes the thermoacoustic transduction medium, adapts easily and continuously to the sample shape. It is therefore not necessary to deal with thermal contact issues that would undoubtedly provoke signal drifts during long-term experiments. However, it is crucial to properly control the cavity sealing state, as changes may induce variations in the cavity thermoacoustic response. 

Finally, replacing an optical sensor by a microphone facilitates instrument integration. It is not necessary to collect the optical beam after interaction with the sample, which eliminates the need for precise and tedious optical alignment. This feature makes the solution more suitable for embedded applications.

For all these reasons, photothermal spectroscopy, and more specifically IPAS, appears to be a promising solution for performing long-term IR spectroscopy on plant stems. The richness of the photoacoustic signal should help capture the complexity of the stem signal, while the air cavity should allow dealing with a large range of plant morphologies. Nonetheless, it might be necessary to develop specific sealing solutions for carefully closing the PA cavity, and physiological models to interpret data. In the literature, photoacoustic spectroscopy has already been used on plants \cite{pereira_opc_1992, mesquita_open_2006} but in another spectral range (near-infrared) and adapted to leaves rather than stems.

\subsection{Mathematical description of the photoacoustic measurement presented in this manuscript}

Based on the previous explanations, a few notations and concepts should be introduced before proceeding.

The PA spectroscopy experiments performed within the scope of this manuscript consist in repeatedly acquiring a pattern of individual measurements over time. Each measurement $s ~\mathrm{[V]}$ is an acoustic recording realized at a fixed modulation frequency $f$ and optical wavelength $\lambda~$. The pattern of $f$ and $\lambda$ scans depends on the characteristics of the instruments and the experimental goal. The photothermal physical process and its associated detection chain both have a linear response. As a consequence, the measurement $s$ can be reduced to the extraction of its Fourier coefficients at $f$. In the following, $s$ is a complex number characterized by its modulus $\rho~\mathrm{[V]} = \lvert s \rvert$ designating the PA effect amplitude, and phase $\phi~\mathrm{[rad]} = \mathrm{arg} (s)$, representing its phase shift with respect to the optical source modulation. $s$ can be decomposed into the successive physical processes involved:

\begin{itemize}
\item $i~\mathrm{[W]}$: The optical source power at $\lambda$ and $f$. The optical source is chosen as the phase reference (its phase is null). Experimentally, the instrument responsible for the optical emission is usually adjusted to keep the optical power constant independently of $\lambda$ and $f$.
\item $R_{PT}~\mathrm{[K.m^2.W^{-1}]}$: The sample photothermal response at the air cavity contact. This is where the sample spectroscopy takes place. $R_{PT}$ depends on the sample optical properties $\mathbf{p}_{opt}^{stem}$. Here, $\mathbf{p}$ designates a bundle of parameters that are sufficient to describe the optical absorption coefficient $\beta$ scalar field within the sample. If the sample is homogeneous, $\mathbf{p}$ is a single scalar value, if it consists of a succession of layers, it may contain each layer absorption coefficient and thickness, etc. $R_{PT}$ also depends on the sample and the cavity gas thermomechanical parameters, respectively $\mathbf{p}_{thm}^{stem}$ and $\mathbf{p}_{thm}^{gas}$. Just like $\mathbf{p}_{opt}$, $\mathbf{p}_{thm}$ contains sufficient information to describe the gas or the sample thermomechanical properties (thermal conductivity, thermal capacity, density) spatial distribution. Finally, $\mathbf{p}_{opt}^{stem}$ depends on $\lambda$. A mathematical description of $R_{PT}$ has already been written for homogeneous \cite{rosencwaig_theory_1976} and stratified samples \cite{hu_generalized_1999}.
\item $R_{AC}~\mathrm{[Pa.K^{-1}.m^{-2}]}$: The PA cavity thermoacoustic response. This response depends on the cavity geometrical and environmental parameters (respectively $\mathbf{p}_{geo}^{gas}$ and $\mathbf{p}_{env}^{gas}$). $\mathbf{p}_{geo}^{gas}$ at least contains the cavity volume $V~\mathrm{[m^3]}$, and $\mathbf{p}_{env}^{gas}$ the average gas temperature $T~\mathrm{[K]}$ and pressure $P~\mathrm{[Pa]}$. $R_{AC}$ is sensitive to $f$, and if the gas is transparent to the optical source it should not depend on $\lambda$.
\item $R_{EL}~\mathrm{[V.Pa^{-1}]}$: The pressure signal conversion to a digitized voltage. This element comprises the microphone and the electronic readout responses. $R_{EL}$ depends on $f$, but should be designed to provide a response as constant as possible. It may also depend on environmental conditions if the measurement is performed in a highly fluctuating environment (which is not the case within the scope of this manuscript since the measurements presented were performed in a laboratory).
\end{itemize}
All in all, $s$ reads:

\begin{equation}
s (f, \lambda, \mathbf{p}) = R_{EL} (f) \times R_{AC} (f, \mathbf{p}_{geo}^{gas}, \mathbf{p}_{env}^{gas}) \times R_{PT} (\lambda, f, \mathbf{p}_{thm}^{gas}, \mathbf{p}_{opt}^{stem}, \mathbf{p}_{thm}^{stem}) \times i ~~~\mathrm{.}
\label{eq:s_definition}
\end{equation}

This representation of the signal $s$ is interesting as it allows visualizing which properties impact which parts of the detection chain. It is further discussed in Sec.~\ref{sec:results}.

\section{Materials and Methods}
\label{sec:materials}

A long-term monitoring experiment was conducted on a plant sample. The plant studied, the experimental setup, the measurement protocol, and the data processing procedure are described below.

\subsection{Plant sample}

The monitoring experiment was conducted on the petiole of an \textit{Oxalis triangularis}. Petioles have a vascular organization similar to that of stems. In the following, the two terms are used interchangeably when referring to the monitored area. This species was selected because it has thin petioles (approximately $1.5 ~\mathrm{mm}$ diameter) containing several vascular bundles, providing a suitable target for measurement within vascular tissues. It also exhibits pronounced circadian-driven nyctinastic leaf movements, which are genetically regulated through changes in turgor pressure and primary metabolism \cite{tuo_unraveling_2025}. 

\subsection{A photoacoustic setup adapted to plant monitoring}
\label{subsec:pa_for_stem}

Fig.~\ref{fig:oxalis_and_setup} presents a scheme of the experimental setup, which elements are successively described below, following the course of a typical measurement process.

The IR source is a Quantum Cascade Laser (QCL) (purchased from Alpes Laser, mounted and collimated in an HHL package) whose emission wavelength is fixed at $ \lambda = 9.65~\mathrm{\mu m}$. This wavelength corresponds to glucose main MIR absorption peak, and is situated within a water transparency region. The QCL is powered by a sine-modulated with a DC offset current source (THORLABS ITC-4001), so that the optical power at the sample level is sinusoidally modulated with a mean value and amplitude of $5~\mathrm{mW}$. The sine-wave frequency $f$ is adjusted throughout the experiment according to the chosen protocol. Before reaching the sample, part of the optical power is redirected to an optical powermeter (THORLABS S401C) using a beamsplitter, for a continuous control of the QCL stability along the experiment.

After the beamsplitter, the IR beam has a diameter of a few millimeters. A $40~\mathrm{mm}$ converging lens is then used to focus the beam on the sample. By reducing the beam spot size at the sample level, the scanned surface is roughly planar, thereby easing signal interpretation. Secondly, it limits the illumination of the PA cavity walls, which could generate an unwanted signal that is not specific to plants.

\begin{figure}[htbp]
\centering
\includegraphics[width=0.75\linewidth]{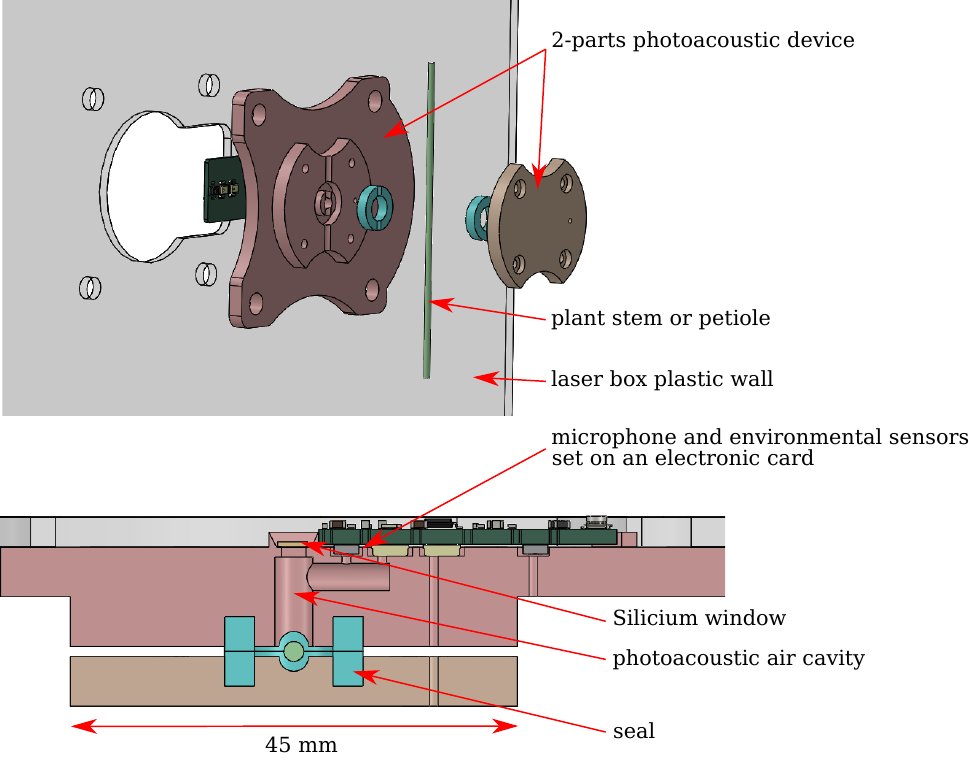}
\caption{3D views of the photoacoustic device. The top figure shows an exploded view of the different components of the device. The plant stem is represented as a vertical green cylinder. The bottom figure shows a cross-sectional view of the assembled device. It illustrates the photoacoustic cavity, the plant stem enclosed within it, and how it is closed by the silicon window and electronic sensors. A picture of the setup is also shown in Fig.\ref{fig:oxalis_and_setup}.}
\label{fig:setup_cao}
\end{figure}

The PA cavity consists of two plastic parts fabricated using a SLA 3D printer (see Fig.~\ref{fig:setup_cao}). The first part is fixed and aligned with the IR beam, while the second is screwed onto the first when the experiment begins. Once the two parts are attached, the PA cavity is closed and the plant stem is enclosed within the cavity. The cavity airtightness stability is critical, as any leak may modify the acoustic cavity response and introduce a drift into the spectroscopy signal. To ensure airtightness, a custom seal matching the plant stem diameter (the 2 blue parts on Fig.~\ref{fig:setup_cao}) was designed and fabricated. The IR beam passes through the PA cavity via an anti-reflective coated silicon window fixed on the plastic first part with UV glue. The acoustic signal is recorded with a MEMS analogical microphone (KNOWLES~SPU410HR5H-PB), mounted on a homemade electronic board, and digitized using a conventional data logger (NI-DAQ~9215). This electronic board also includes temperature and relative humidity sensors (SENSIRION SHT35A-DIS-B). It is glued to the plastic fixed part, and connected to the PA cavity through dedicated air channels.

The PA cavity total air volume is about $300~\mathrm{mm^3}$ of which circa $15~\mathrm{mm^3}$ is occupied by the plant stem. It is assumed that the sealing quality is good enough to prevent pressure fluctuations associated with the spectroscopy measurement from escaping the cavity, while remaining sufficiently permeable to allow pressure and humidity to equilibrate with the outside. This assumption is discussed further in Sec.~\ref{sec:results}.

The instrumental setup is located in a laboratory environment, where temperature and humidity fluctuations are minimal. The room has windows, so the plant receives indirect natural light throughout the experiment. The environmental parameters of the room (temperature, relative humidity, pressure, illumination) are recorded using a basic environmental station (ARDUINO MKR ENV SHIELD), which is also programmed to record the environmental parameters measured inside the PA cavity. This setup is not typical of plant biology experiments. For example, a controlled lighting system and a more standaradized plant sample would be preferable. However, this experiment is sufficient to support the conclusions drawn within the scope of this manuscript.

Fig. \ref{fig:oxalis_and_setup} shows a picture of the plant during a measurement. 

\subsection{Acquisition protocol and data processing}
\label{subsec:acquisition_protocol}

Two experiments are presented in this manuscript. A first control experiment was conducted without a sample in the PA cavity for 17 hours to monitor possible instrumental drifts and effects related to the setup and environmental conditions. The second experiment was conducted with a plant sample in the PA cavity for 3.5 days. For both experiments, data were recorded every $45~\mathrm{s}$. Each acquisition consisted in recording all environmental parameters and three consecutive photoacoustic measurements. The three photoacoustic measurements differed in the laser intensity modulation frequency $f$, which was successively set to $[157, 282, 507]~\mathrm{Hz}$. These frequencies were chosen to provide different thermal diffusion lengths, as explained in Sec.~\ref{subsec:advantages_and_drawbacks}. Each acoustic recording was $1~\mathrm{s}$ long. Thus, each acquisition comprised six photoacoustic quantities (modulus and phase for each modulation frequency), and six environmental parameters (temperature and relative humidity inside the photoacoustic cavity, temperature, relative humidity, illumination and pressure outside the photoacoustic cavity). Once digitized, the signal was demodulated at the optical excitation modulation frequency using a numerical lock-in amplifier. The spectroscopy data time series were finally smoothed with a $2$-hour-wide median filter (the choice for the time averaging window width is discussed in \ref{ann:allan_study}). 

\section{Results and Discussion}
\label{sec:results}

In this section, measurements from a control (Fig.~\ref{fig:results_reference}) and a plant (Fig.~\ref{fig:results_plant}) experiments are presented. For each experiment, observations on the data are first made, followed by a discussion of possible interpretations. The control experiment provides insights into instrumental drift in the setup, thereby helping to distinguish it from the physiological response observed in the plant experiment. Only the most relevant information is presented in Figs.~\ref{fig:results_reference},\ref{fig:results_plant}, while the complete environmental data and the cross-plots between spectroscopic and environmental data are provided in \ref{ann:env_plots} and \ref{ann:cross_plots}.

\subsection{The control experiment}
\label{subsec:ref_experiment}


\subsubsection*{Observations}

\begin{figure}[htbp]
\centering
\includegraphics[width=1.00\linewidth]{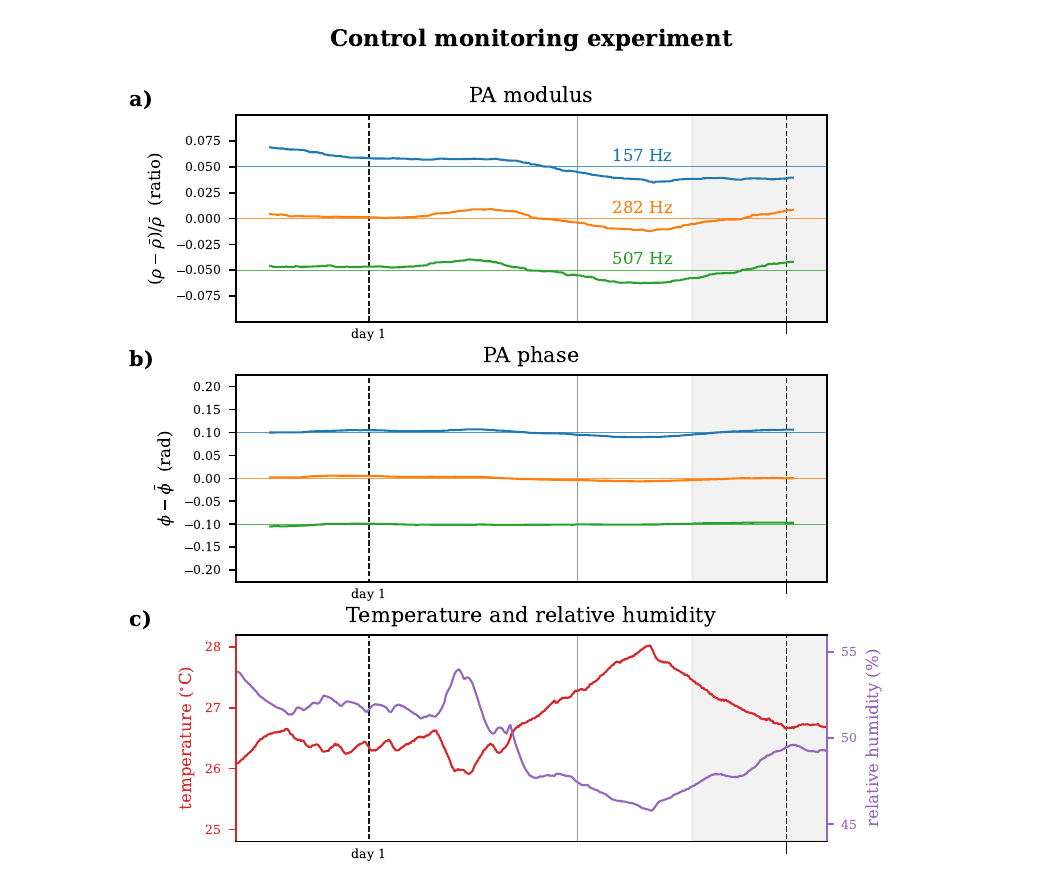}
\caption{Control experiment spectroscopy and environmental time series. Subfigures \textit{a)} and \textit{b)} show the modulus and the phase of the PA data, respectively. The modulus is centered and normalized by its average value $\bar{\rho}$. The phase is centered by its average value $\bar{\phi}$. For visualization purposes, the datasets associated to the different modulation frequencies are offseted to arbitrary values (represented by horizontal lines in the same colour as the corresponding datasets). The vertical scales in Fig.~\ref{fig:results_reference} and Fig.~\ref{fig:results_plant} are identical to facilitate direct comparison. Subfigure \textit{c)} shows the PA cavity temperature and relative humidity. Based on the illumination data, a gray area is used to indicate periods of low light periods, which corresponds here to night. Vertical black dashed lines indicate noon and midnight.}
\label{fig:results_reference}
\end{figure}

While the control experiment is short ($17$ hours), it still encompasses characteristic daily environmental variations experienced by the system (Fig.~\ref{fig:results_reference}c). The temperature inside the PA cavity oscillates between $26$ and $28~\mathrm{^{\circ}C}$, and is strongly anticorrelated with relative humidity. The external temperature and relative humidity mimic those inside the PA cavity (see \ref{ann:env_plots}). The external ambient pressure is roughly anticorrelated with the temperature. It varies over a range of $0.30~\mathrm{kPa}$ around an average value of $99.65~\mathrm{kPa}$. Since the PA cavity is expected not to be perfectly airtight, its internal average pressure should follow the external one.

The spectroscopy signal (Fig.~\ref{fig:results_reference}a-b) exhibits fluctuations with amplitudes of approximately $2~\%$ at $157~\mathrm{Hz}$ and $1~\%$ at $[282, 507]~\mathrm{Hz}$ for the PA modulus, and $5~\mathrm{mrad}$ for the phase. The modulus is anticorrelated with the temperature, with a slope of approximately $-1~\mathrm{\%.^{\circ}C^{-1}}$ (see \ref{ann:cross_plots}), independent from the modulation frequency. The phase is also anticorrelated with the temperature, but its slope decreases as the modulation frequency increases. It decreases from $-7.5~\mathrm{mrad.^{\circ}C^{-1}}$ at $157~\mathrm{Hz}$ to an undetectable value at $507~\mathrm{Hz}$. Finally, the $157~\mathrm{Hz}$ modulus appears to exhibit a global decreasing trend.

\subsubsection*{Impact of the cavity gas average temperature and pressure}

For the control experiment, the only source of variability in the signal $s$ should be the cavity response $R_{AC}$ (see Eq.~\ref{eq:s_definition}). Indeed, there is no living sample in the cavity (the sample is the plastic walls of the PA cavity), so the photothermal response $R_{PT}$ should remain constant. The electronic response $R_{EL}$ should not be affected by such weak environmental fluctuations. The most common and simplest mathematical expression for $R_{AC}$ is the \textit{thermal piston} model \cite{hu_generalized_1999}. $s$ reads:

\begin{equation}
s = R_{EL} \times \underbrace{\frac{P \mu_{gas} \mathrm{e}^{-j \pi / 4}}{\sqrt{2} T V}}_{R_{AC}}  \times R_{TP} \times i ~~~\mathrm{,}
\label{eq:hu_1}
\end{equation}
where $P~\mathrm{[Pa]}$, $T~\mathrm{[K]}$, and $V~\mathrm{[m^3]}$ correspond to the average pressure and temperature of the gas inside the cavity and to the cavity volume, respectively. The relative sensitivity of $s$ to these parameters reads:

\begin{equation}
\frac{\mathrm{d s}}{s} = - \frac{\mathrm{d} T}{T} + \frac{\mathrm{d} P}{P} - \frac{\mathrm{d} V}{V} ~~~\mathrm{.}
\label{eq:s_sensitivity_1}
\end{equation}

For the experiment atmosphere, $T \approx 300~\mathrm{K}$, and $P \approx 10^5 ~\mathrm{Pa}$. Assuming that the cavity volume remains constant, Eq.~\ref{eq:s_sensitivity_1} simplifies to:

\begin{equation}
\frac{\mathrm{d s}}{s} \approx - 1/300 \times \mathrm{d} T ~\mathrm{[K]} + 1/100 \times \mathrm{d} P ~\mathrm{[kPa]}  ~~~\mathrm{.}
\label{eq:s_sensitivity_2}
\end{equation}

In other words, a $3~\mathrm{^{\circ}C}$ temperature change or a $1~\mathrm{kPa}$ pressure change is sufficient to generate a $\sim 1~\%$ relative change in $s$. In addition, these parameters should only affect the modulus $\rho$ of $s$, and not the phase $\phi$.

\begin{figure}[htbp]
\centering
\includegraphics[width=1.00\linewidth]{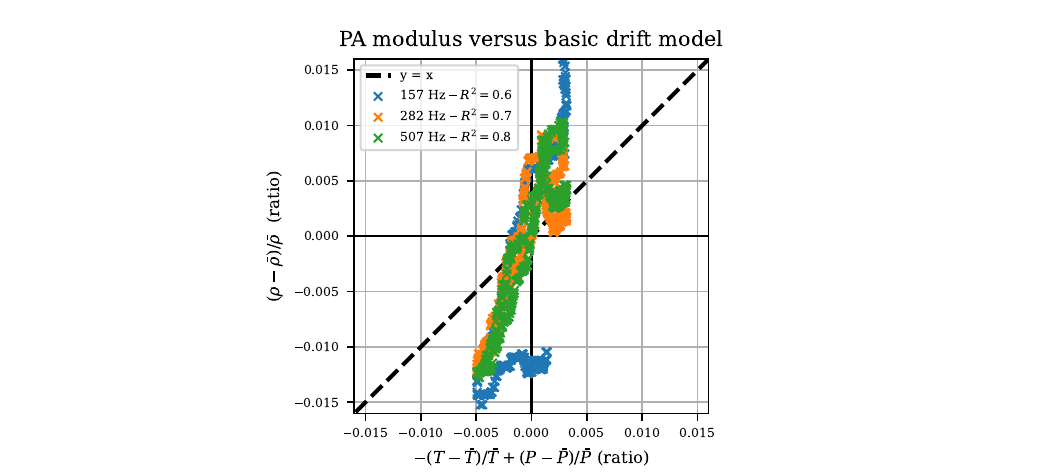}
\caption{Cross-plot of the control experiment spectroscopy data and a basic model based on environmental parameters. The relative PA modulus $(\rho - \bar{\rho}) / \bar{\rho}$ at the three modulation frequencies is plotted as a function of the model given by Eq.~\ref{eq:s_sensitivity_1}. If the model accurately described the spectroscopy data, the measurement points would lie on the black dashed line $y = x$.}
\label{fig:env_basic_model}
\end{figure}

This model is consistent with many observations: the modulus is anticorrelated with temperature variations and correlated with pressure fluctuations. However, it explains only about one-third of the observed drift amplitude (see Fig.~\ref{fig:env_basic_model}), and does not explain the decreasing trend of the $157~\mathrm{Hz}$ modulus, nor the weak dependence of the phase on $T$ and $P$.

\subsubsection*{Impact of gas leaks between the PA cavity and the outside}

Next, the model given in Eq.~\ref{eq:hu_1} assumes that the cavity sealing sufficiently tight to prevent pressure variation associated with the photothermal effect from leaking out of the cavity. Nevertheless, the sealing is not perfect, which implies that the cavity pressure, temperature, and humidity eventually equilibrate with the external environment. The cavity sealing allows slowly varying pressure exchanges while preventing fast ones. In order to contain the PA effect, the filter cutoff frequency $f_{leak}~\mathrm{[Hz]}$ must remain much smaller than the optical modulation frequency. The porrer the sealing, the higher $f_{leak}$ becomes. A common modeling approach \cite{lauwers_all-optical_2020} is to model the leakage contribution as a first-order high-pass filter. $R_{AC}$ then reads:

\begin{equation}
R_{AC} = \underbrace{\frac{j f / f_{leak}}{1 + j f / f_{leak}}}_{\mathrm{leak~contribution}} \times \underbrace{\frac{P \mu_{gas} \mathrm{e}^{-j \pi / 4}}{\sqrt{2} T V}}_{\mathrm{hermetic~cavity}} ~~~\mathrm{,}
\label{eq:improved_R_AC}
\end{equation}
with:

\begin{equation}
f_{leak} =  \frac{1}{2 \pi} \times \frac{P d^4}{12 \gamma V L \nu} \approx \frac{P d^4}{106 V L \nu} ~~~\mathrm{,}
\label{eq:f_leak}
\end{equation}
where $\nu~\mathrm{[Pa.s]}$ and $\gamma~\mathrm{[no~unit]}$ denote the dynamic viscosity of the cavity gas and ratio of specific heats, respectively. $d~\mathrm{[m]}$ and $L~\mathrm{[m]}$ represent the leak conduit diameter and length, respectively. This model assumes that the leak is a cylinder conduit connecting the cavity to the outside. For example, if $d = 100~\mathrm{\mu m}$ and $L = 2 ~\mathrm{mm}$, which is a realistic situation for the experiment presented in this manuscript, $f_{leak} \approx 7 ~\mathrm{Hz}$. The sealing quality, or at least control, is therefore a fundamental aspect of the acoustic response stability, as $f_{leak}$ may approach $f$. It is possible that the slow decrease observed at $157~\mathrm{Hz}$ is caused by a progressive degradation of the sealing quality. Such a scenario would increase $f_{leak}$ and consequently affect the $157~\mathrm{Hz}$ measurements before those at higher frequencies. However, according to Eq.~\ref{eq:improved_R_AC}, it should also induce a strong phase shift, which is not observed.

\subsubsection*{Limits of the thermal piston hypothesis}

Eq.~\ref{eq:hu_1} provides a valid description of $R_{AC}$ within the limits of the so-called \textit{thermal piston} regime. It assumes that the acoustic wavelength remains large with respect to the characteristic dimensions of the cavity. In other words, the cavity is operated in a frequency regime that stays below its first acoustic resonance, such that the acoustic pressure remains spatially approximately constant within the cavity. As a consequence, the \textit{thermal piston} makes the PA measurement independent of propagation properties of acoustic waves in the gas. This is why, in this regime, $R_{AC}$ phase is independent of $f$. 

On the other hand, a frequency-dependent phase may appear in other regimes, for example, if the optical modulation frequency approaches the cavity acoustic resonances. In the presented experiment, the cavity largest dimension is $\sim 20 ~\mathrm{mm}$ (see Fig.~\ref{fig:setup_cao}). This implies that the first acoustic resonance should occur around $\sim 10 ~\mathrm{kHz}$, which is above the highest modulation frequency ($507~\mathrm{Hz}$). Nonetheless, deviation from the \textit{thermal piston} regime may contribute to the weak temperature dependence observed in the phase and may explain why this dependence tends to disappear as $f$ decreases.

\subsubsection*{Partial conclusion for the control experiment}

The control experiment signal fluctuations should only be caused by variations in the PA cavity response $R_{AC}$. The fluctuations represent $1-2~\%$ for the modulus, and $5~\mathrm{mrad}$ for the phase. The internal temperature and pressure of the cavity are sufficient to model empirially these fluctuations within the limits of signal noise, with the exception of the modulus at the lowest modulation frequency. This latter phenomenon could be explained by the gradual development of a leak between the cavity and the outside.

Although the modeling approaches described in the previous subsections provide some useful insights, they are not sufficiently developed to fully explain the observed data. In the future, it would be worthwile to repeat this experiment with greater environmental fluctuations and a wider range of modulation frequencies. The data collected should lead to a better understanding of the physical phenomena involved and help develop of a more comprehensive model of the cavity response.

Last but not least, $R_{AC}$ does not depend on $\lambda$ as long as the optical absorption of the gas remains sufficiently weak. Therefore, a straightforward solution to compensate for the cavity response and, consequently, its drifts would be to repeat the acquisition at several $\lambda$, and normalize $s$ to a reference wavelength.

\subsection{The plant experiment}
\label{subsec:stem_experiment}

\begin{figure}[htbp]
\centering
\includegraphics[width=1.00\linewidth]{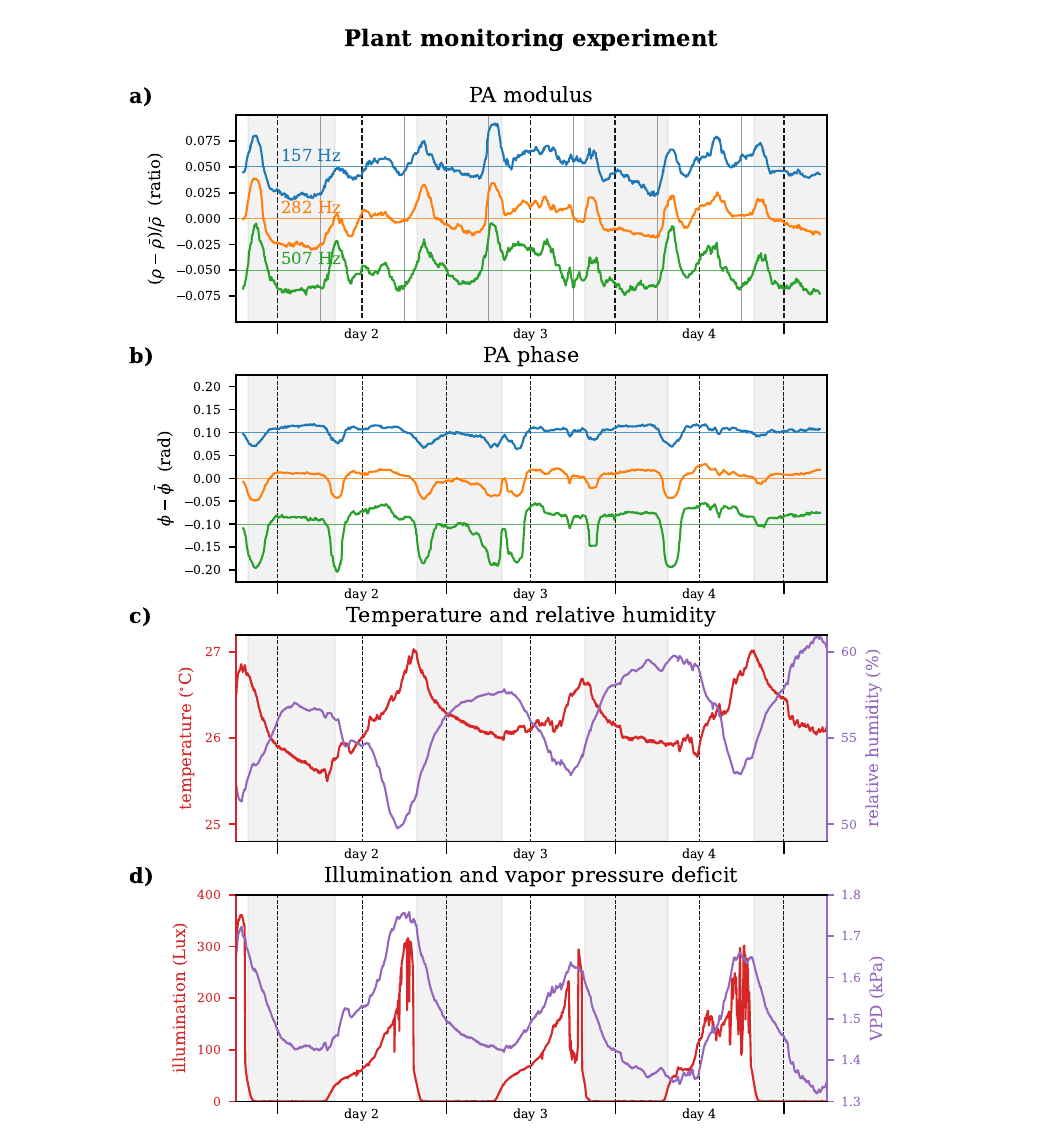}
\caption{Control experiment spectroscopy and environmental time series. Subfigures \textit{a)} and \textit{b)} show the modulus and the phase of the PA data, respectively. The modulus is centered and normalized by its average value $\bar{\rho}$. The phase is centered by its average value $\bar{\phi}$. For visualization purposes, the datasets associated to the different modulation frequencies are offseted to arbitrary values (represented by horizontal lines in the same colour as the corresponding datasets). The vertical scales in Fig.~\ref{fig:results_reference} and Fig.~\ref{fig:results_plant} are identical to facilitate direct comparison. Subfigure \textit{c)} shows the PA cavity temperature and relative humidity. Subfigure \textit{d)} shows illumination and vapor pressure deficit (VPD). Based on the illumination data, a gray area is used to indicate periods of low light periods, which corresponds here to night. Vertical black dashed lines indicate noon and midnight.}
\label{fig:results_plant}
\end{figure}

\subsubsection*{Observations}

\begin{figure}[htbp]
\centering
\includegraphics[width=1.00\linewidth]{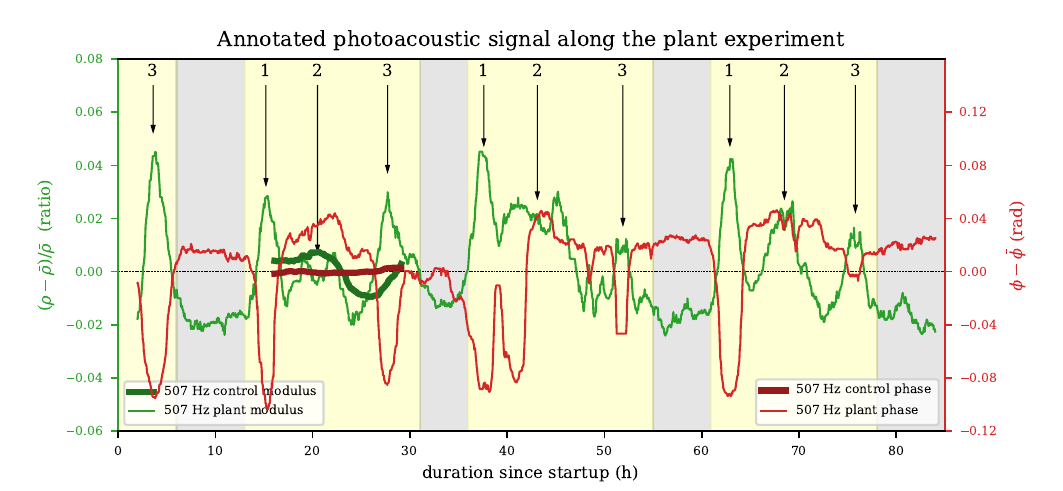}
\caption{Plant experiment spectroscopy signal at $\mathrm{507~\mathrm{Hz}}$ with annotations. The relative modulus $(\rho - \bar{\rho})/\bar{\rho}$ and phase $\phi - \bar{\phi}$ are superimposed on the same graph for the plant (solid thin lines) and the control (solid thick lines) experiments. The control experiment is time-shifted to the plant experiment second day for readability. Yellow and black regions correspond to the area defined as day and night in Sec.~\ref{subsec:stem_experiment}. The three peaks observed each day in the modulus are annotated.}
\label{fig:results_plant_annotated}
\end{figure}

During the three-and-a-half-day acquisition period presented in Figs.~\ref{fig:results_plant},\ref{fig:results_plant_annotated} (and in \ref{ann:env_plots} and \ref{ann:cross_plots} for complementary data), the environmental parameters follow a pattern very similar to that observed in the control experiment. Diurnal variations occur both inside and outside the cavity, with temperature fluctuations between $25.5$ and $27~\mathrm{^{\circ}C}$ and relative humidity fluctuations between $50$ and $60~\mathrm{\%}$, resulting in VPD (Vapor Pressure Deficit) values between $1.3$ and $1.8~\mathrm{kPa}$. Temperature and relative humidity are anticorrelated. External pressure is anticorrelated with temperature, and varies between $99.2$ and $99.6~\mathrm{kPa}$.  

More remarkably, the photothermal signal, both in modulus and phase, also exhibits a strong diurnal pattern. Each day, the modulus can be divided into two consecutive regimes. The first starts at sunrise, and the second a few hours after sunset (hereafter referred to as “day” and “night”, respectively). On average, the modulus was higher during the day than during the night. More precisely, during the day, it can be separated into three consecutive positive peaks lasting between $4$ to $6~\mathrm{h}$. During the night, the modulus slowly decreases. The amplitude of these fluctuations is typically about $3~\%$. The overall pattern is repeated at all modulation frequencies, with small frequency-dependent variations. At the beginning and end of the day, the phase exhibits two negative peaks that appear to be anticorrelated with the first and third modulus peaks, respectively. Their amplitudes range from $\sim 20~\mathrm{mrad}$ at $157~\mathrm{Hz}$ to $\sim 100~\mathrm{mrad}$ at $507~\mathrm{Hz}$. The second modulus peak may be positively correlated with the phase, but with a weaker amplitude of a few tenths of $\mathrm{mrad}$. Those observations are summarized in Fig.~\ref{fig:results_plant_annotated}.

\subsubsection*{Impact of instrumental drifts}

In contrast to the control experiment, there is no simple way to explain the spectroscopy signal using the environmental sensors data (see \ref{ann:cross_plots} for cross-plots). Moreover, the amplitudes of the signal fluctuations, particularly thos observed in the phase, are larger than those observed during the control experiment. This supports the conclusion that instrumental drifts account for only part of the measured signal. To visualize this difference, the datasets from the plant and control experiments are superimposed in Fig.~\ref{fig:results_plant_annotated}.

\subsubsection*{Impact of the IR source wavelength}

During the experiment, the optical source was not tunable, restricting the analysis to a single wavelength $\lambda = 9.65~\mathrm{\mu m}$. This wavelength corresponds to the main glucose absorption peak in the MIR spectral range and, more generally, to a region of strong absorption by carbohydrates. It also lies within a so-called transparency region (between $7$ and $10~\mathrm{\mu m}$) where water absorption is weak and relatively flat.

However, even at $9.65~\mathrm{\mu m}$, water absorption remains predominant in biological samples because of its high concentration. Consequently, the presented protocol does not allow an unambiguous attribution of the PA signal fluctuations to specific molecules. Additional information on the physiological and biochemical composition of the plant, together with spectroscopic measurements acquired at other wavelengths, would be necessary to enable such an identification.

\subsubsection*{Impact of the stem expansion on the cavity volume}

It is known that plant stems undergo daily expansion due to water circulation cycles within the xylem \cite{sevanto_effects_2011}. The stem expands at the beginning of the day as photosynthesis activity starts, and contracts at night.

As this effect may have an impact on the stem physiological composition (e.g., an increase in water content, an increase in the thickness of the different living tissues), it should also influence the acoustic cavity response $R_{AC}$ by modifying its volume $V$ (see Eq.~\ref{eq:s_sensitivity_1}). For example, assuming a conservative stem diameter expansion of $5~\%$, the resulting relative variation in the modulus during the day of $- \mathrm{d} V / V \approx 0.05 \times 15 / 300 \approx 0.25~\%$. Stem expansion effects should therefore account for only a minor part of the signal variations during the day.

\subsubsection*{A simple model for the modulus-phase anticorrelation}

The strong anticorrelation between the modulus and the phase during the first and the last peaks of the day is particularly interesting, as there is no obvious way to attribute it to the cavity response $R_{AC}$. However, a possible explanation may be found in the plant photothermal response $R_{PT}$.

For a homogeneous sample, it can be shown (see \ref{ann:pt_sensitivity}) that both the PA modulus and phase are positively correlated with the optical absorption coefficient $\beta$. A change of a few percent in $\beta$ should induce a few percent increase in $\rho$, and a few $\mathrm{mrad}$ increase in $\phi$. This may explain what occurs during the second peak of the day.

While the modulus is strongly related to optical absorption, the phase is primarily influenced by the sample structure when the latter is heterogeneous. For example, consider a two-layer sample consisting of a non-absorbing first layer and an absorbing second layer. Both layers are thermally conductive. If the thickness of the first layer increases by $\delta l~\mathrm{[m]}$, the heat coming from the second layer should take longer to reach the surface, thereby decreasing the signal phase. A reasonable order-of-magnitude for the resulting delay is $\delta t ~[\mathrm{s}] \approx - \delta l^2 / \alpha$, where $\alpha~[m^2. s^{-1}$] is the sample thermal diffusivity, and $\delta \phi = 2 \pi f \delta t$. For $\delta l = 5 ~\mathrm{\mu m}$ in water at $f = 282~\mathrm{[Hz]}$, this gives $\delta \phi \approx - 75 ~\mathrm{mrad}$. Such a variation is of the same order of magnitude as the observed signal. Moreover, by definition, the amplitude of $\delta \phi$ increases with $f$, which is consistent with the data.

To conclude, a possible explanation for the two daily anticorrelated events could be the following: the IR beam probes two consecutive regions of the stem, and during the peak, the absorption in the second region increases as the thickness of the first layer increases. The system then returns to its initial state. This simple two-layer model has the advantage of being straightforward.

\section{Conclusion}
\label{sec:conclusion}

This work presents MIR indirect photoacoustic spectroscopy as a promising approach for monitoring the physiological activity of vascular plants \textit{in vivo}, non-invasively, and over long durations. The setup included an IR monochromatic laser centered on the main absorption band of glucose which was used to probe the first hundred micrometers of a plant petiole. The spectroscopic response of the sample was then measured through thermoacoustic transduction in a home-made acoustic cavity. 

A 3-day monitoring experiment was conducted. Importantly, the method allowed continuous measurements on a living plant without detectable damage or tissue excision. In contrast, other optical approaches, including laser-based methods, require careful adjustment of parameters such as wavelength, power, and exposure time to avoid altering or damaging the tissue \cite{mekarni_vivo_2024}, \cite{prats-mateu_following_2018}. The acquired photothermal signal exhibited complex oscillations that were not directly correlated with environmental conditions. Instead, the signal was consistently stronger in the morning and weaker at night, at approximately the same times each day, regardless of natural fluctuations of light intensity, temperature, and humidity. Given the probing depth of the system and the anatomy of the petiole, the measured signal is consistent with changes occurring in or around the vascular bundles, driven by circadian rhythms. Comparison with a control experiment provided evidence that part of the signal is biologically related and is not the result of instrumental drift nor sample dilation. However, interpretation of the signal was limited by the complexity of the sample structure and molecular composition, combined with the limited amount of available information. Its precise biochemical and anatomical origins thus remain unclear and require further investigation. 

This exploratory study enabled the identification of several challenges related to non-invasive long-term monitoring, some of which could be addressed by switching to a multichromatic optical source, improving control of environmental conditions during the experiment, and gaining a refined understanding of the plant sample. It also demonstrated the feasibility of the method, and should enable the acquisition of novel continuous spectroscopic datasets capable of capturing plant physiological activity over long durations.

\section*{Declaration of competing interests}

Kevin Jourde and Maeva Doron report financial support was provided by IPCEI Microelectronics and Connectivity. Régis Burlett report financial support was provided by French National Research Agency (ANR) through the project the EXPLOR'AE program HARMONICA (ANR-24-RRII-0003). Kevin Jourde and Héloïse Hannart have patent US20260023063A1 granted to CEA-Leti. Other authors declare that they have no known competing financial interests or personal relationships that could have appeared to influence the work reported in this paper.

\section*{CRediT author statement}

\textbf{Ellyne Liégeois:} Conceptualization, methodology, investigation, writing - original draft. \textbf{Héloïse Hannart:} Conceptualization, methodology, investigation, writing - original draft. \textbf{Jules Skubich:} Conceptualization, investigation. \textbf{Maeva Doron:} Conceptualization, funding acquisition, writing – review and editing. \textbf{Kevin Jourde:} Conceptualization, methodology, investigation, software, funding acquisition, supervision, writing - original draft. \textbf{Johan Decelle:} Supervision, writing – review and editing. \textbf{Yann Bresson:} Conceptualization, writing - original draft. \textbf{Sylvain Delzon:} Conceptualization, supervision. \textbf{Régis Burlett:} Conceptualization, methodology, funding acquisition, supervision, writing - original draft.

\clearpage
\newpage
\appendix

\clearpage
\newpage
\section{Selection of the time averaging window width}
\label{ann:allan_study}

\begin{figure}[htbp]
\centering
\includegraphics[width=0.75\linewidth]{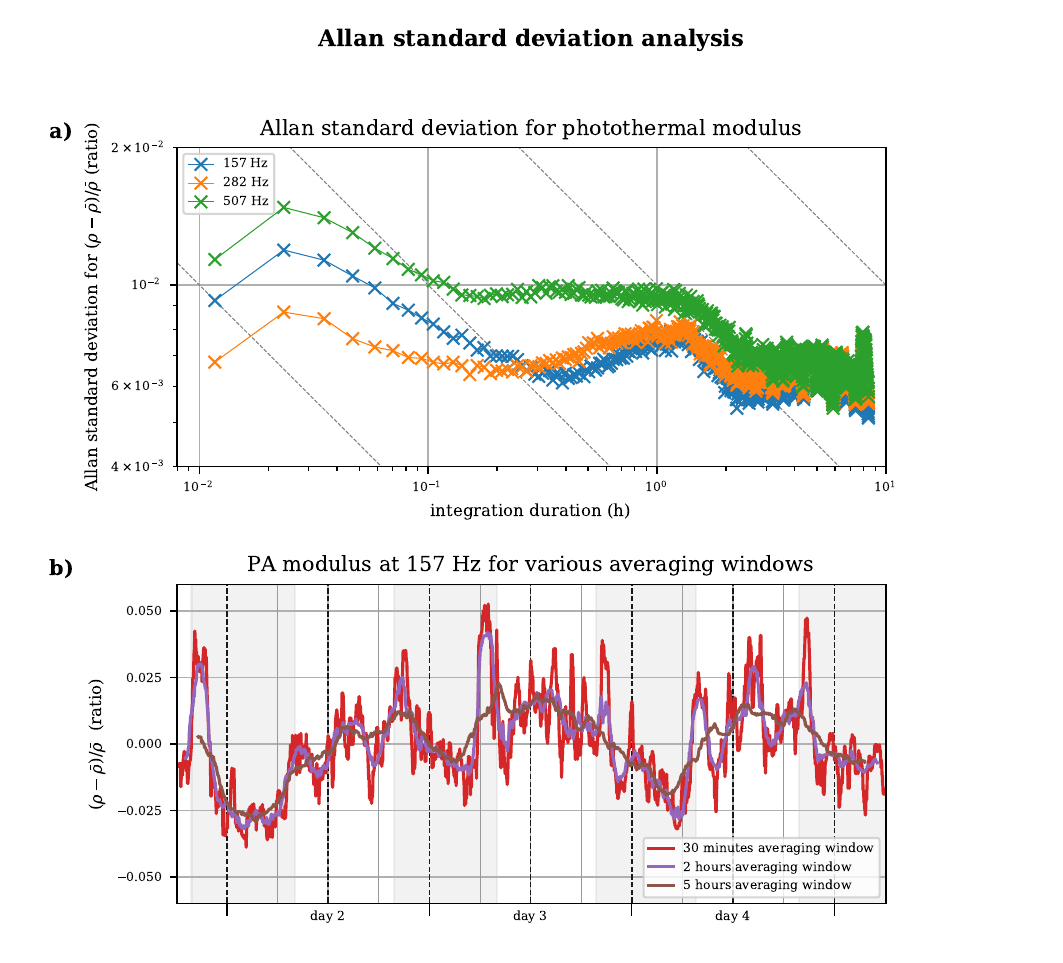}
\caption{Noise analysis of the plant experiment. Subfigure \textit{a)} presents the Allan standard deviation of the relative PA modulus $(\rho - \bar{\rho}) / \bar{\rho}$ for the three modulation frequencies. Black dashed lines correspond to $1/\sqrt{\tau}$ power laws, $\tau~\mathrm{[s]}$ being the integration duration. Subfigure \textit{b)} displays the PA modulus signal as a function of time for $f = 157~\mathrm{Hz}$ and for various averaging windows sizes.}
\label{fig:allan_study}
\end{figure}

Allan analysis of the PA signal from the plant experiment reveals a great amount of information (see Fig.~\ref{fig:allan_study}):

\begin{itemize}
\item The Allan standard deviation starts with an increase while it would be expected that it would fit a $1/\sqrt{\tau}$ power law (where $\tau~\mathrm{[s]}$ is the integration duration) on such a short integration duration. This is explained by the presence of repetitive outliers in the raw signal.
\item It then decreases but still does not fit the $1/\sqrt{\tau}$ power law, suggesting the presence of drifts or a signature of the plant physiological activity.
\item The Allan standard deviation increases again for $\tau \sim 30~\mathrm{min}$, and decreases back for $\tau \sim 2~\mathrm{h}$. This suggests the presence of a repetitive signal with typical $1~\mathrm{h}$ period. This signal is visible when comparing on Fig.~\ref{fig:allan_study}b the curves with a $30~\mathrm{min}$ and a $2~\mathrm{h}$ period. Those fluctuations typically appear during the day, and are also visible on the outside temperature and relative humidity data (see \ref{ann:env_plots}). These fluctuations are probably related to surrounding anthropogenic activity (the room temperature regulation, or surrounding high power technical equipments).
\item The Allan standard deviation finally stabilizes up to $\tau \sim 10~\mathrm{h}$.
\end{itemize}

In order to remove this $1$-hour contribution, it has been decided to smooth the signal with a $2$-hour large averaging window.

\clearpage
\newpage
\section{Experiments environmental data compilation}
\label{ann:env_plots}

\begin{figure}[htbp]
\centering
\includegraphics[width=1.00\linewidth]{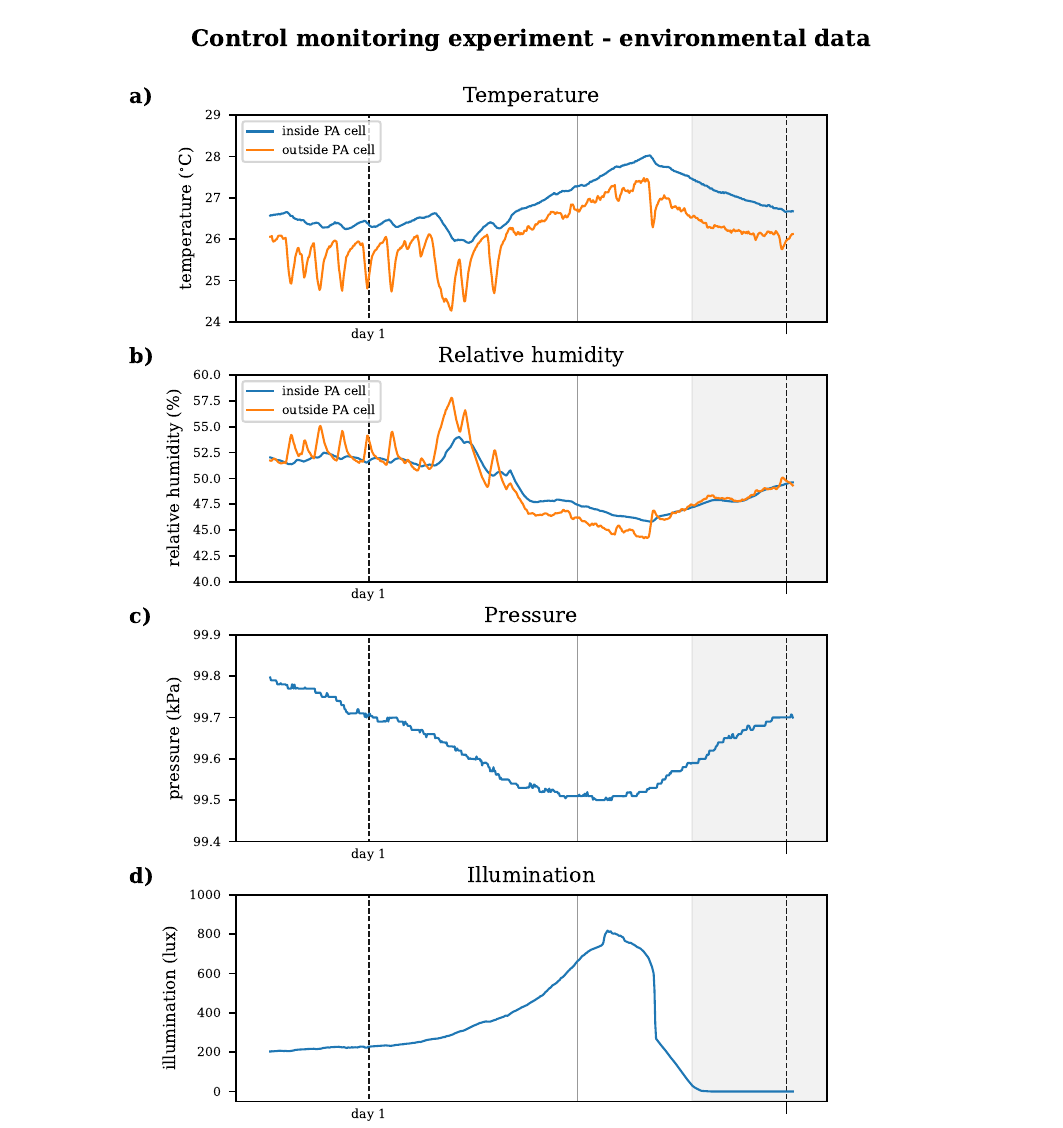}
\caption{Control experiment environmental time series. Subfigures \textit{a)} and \textit{b)} respectively present the temperature and the relative humidity both inside and outside the PA cavity. Subfigure \textit{c)} shows the outside presure, and \textit{d)} the outside illumination. Based on the illumination data, a gray area is drawn for low light periods, here corresponding to night. Vertical black dashed lines represent noon and midnight.}
\label{fig:env_data_ref}
\end{figure}

\begin{figure}[htbp]
\centering
\includegraphics[width=1.00\linewidth]{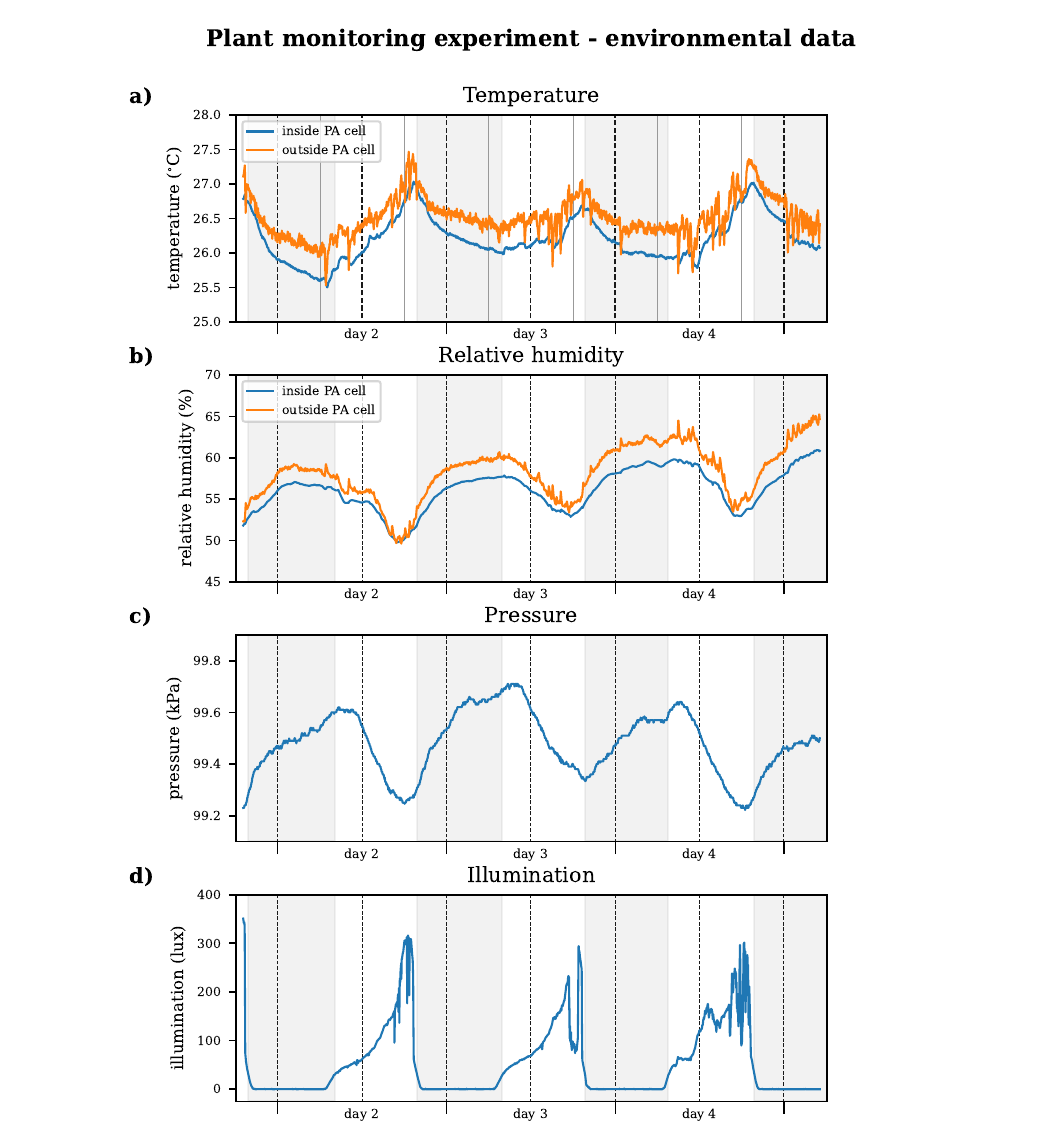}
\caption{Plant experiment environmental time series. Subfigures \textit{a)} and \textit{b)} respectively present the temperature and the relative humidity both inside and outside the PA cavity. Subfigure \textit{c)} show the outside presure, and \textit{d)} the outside illumination. Based on the illumination data, a gray area is drawn for low light periods, here corresponding to night. Vertical black dashed lines represent noon and midnight.}
\label{fig:env_data_plant}
\end{figure}

\clearpage
\newpage
\section{Spectroscopy versus environmental data cross-plots}
\label{ann:cross_plots}

\begin{figure}[htbp]
\centering
\includegraphics[width=1.00\linewidth]{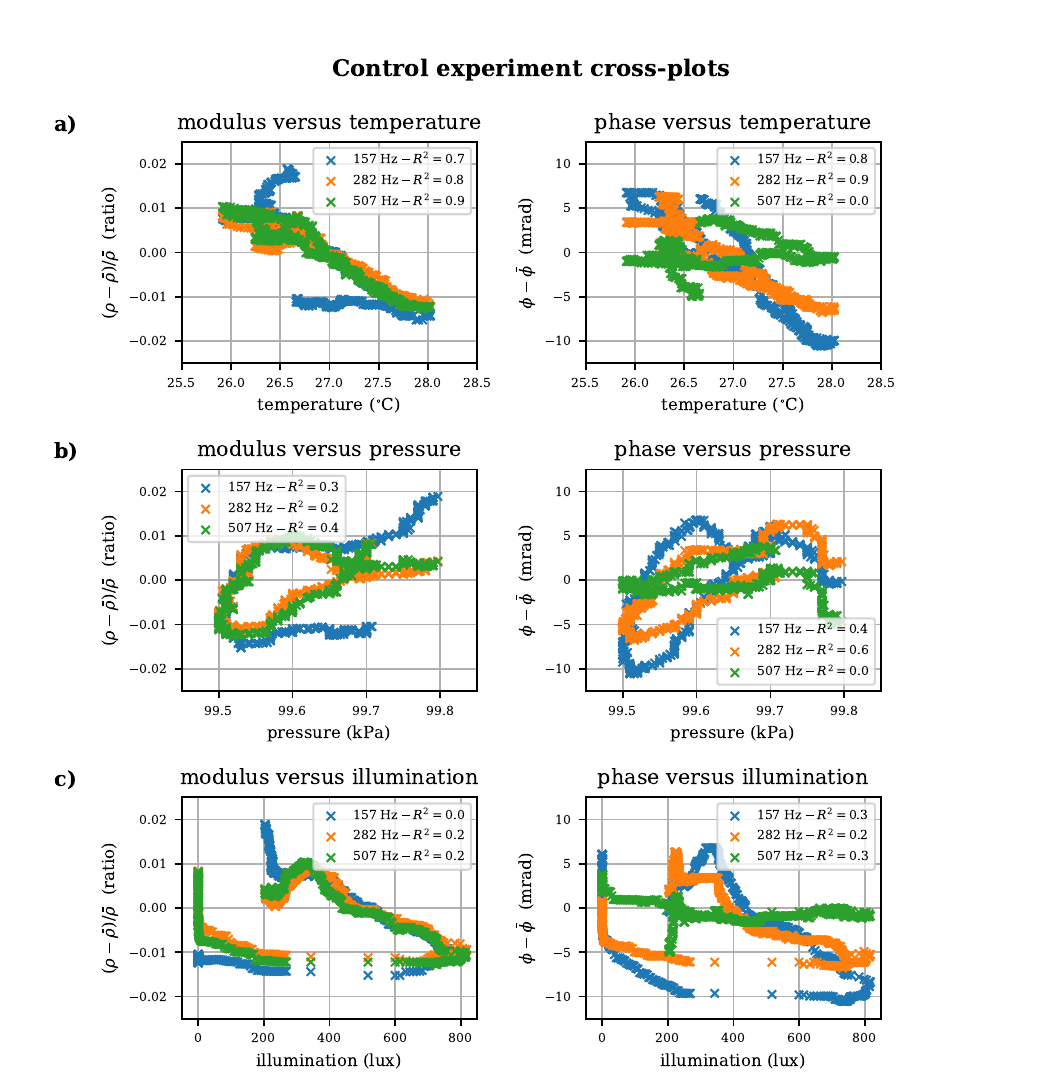}
\caption{Control experiment cross-plots between the spectroscopy (left is modulus, right is phase) and \textit{a)} the temperature data (inside the PA cell), \textit{b)} the pressure data (outside the PA cell), and \textit{c)} the illumination data (outside the PA cell). The modulus is centered and normalized by its average value, while the phase is simply centered by its average value. The different colors are associated to the different modulation frequencies used during the experiments. The legend indicates the linear fit quality $R^2$ for each sub-dataset.}
\label{fig:cross_plot_ref}
\end{figure}

\begin{figure}[htbp]
\centering
\includegraphics[width=1.00\linewidth]{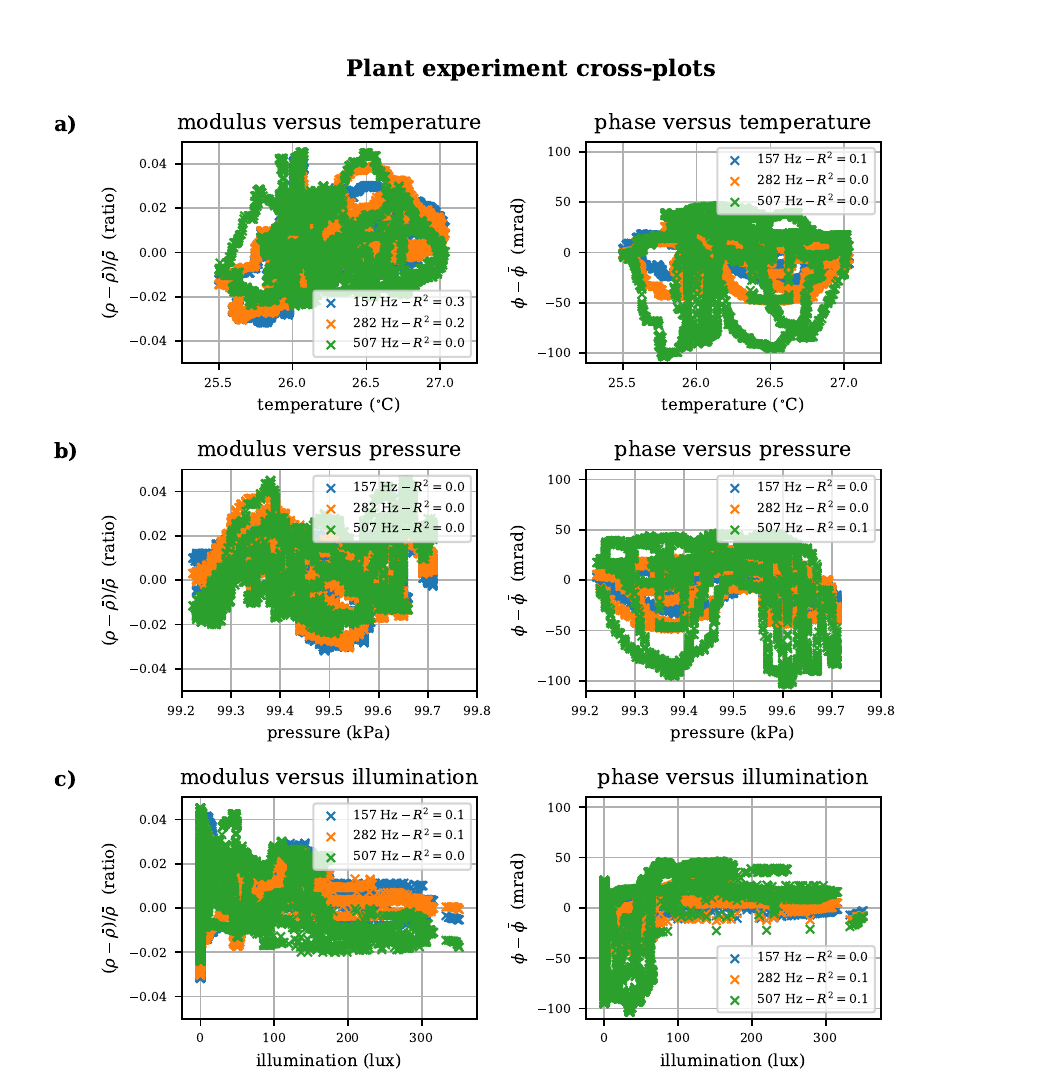}
\caption{Plant experiment cross-plots between the spectroscopy (left is modulus, right is phase) and \textit{a)} the temperature data (inside the PA cell), \textit{b)} the pressure data (outside the PA cell), and \textit{c)} the illumination data (outside the PA cell). The modulus is centered and normalized by its average value, while the phase is simply centered by its average value. The different colors are associated to the different modulation frequencies used during the experiments. The legend indicates the linear fit quality $R^2$ for each sub-dataset.}
\label{fig:cross_plot_plant}
\end{figure}

\clearpage
\newpage
\section{Photothermal response of a homogeneous sample}
\label{ann:pt_sensitivity}

$R_{PT}$ analytical expression for a homogeneous sample reads (adapted from \cite{hu_generalized_1999}):

\begin{equation}
R_{PT} = \frac{\beta}{(\sigma_{stem} + \beta) \times (k_{stem} \sigma_{stem} + k_{gas} \sigma_{gas})} ~~~\mathrm{,}
\label{eq:hu_homogeneous}
\end{equation}
where $k~\mathrm{[W.m^{-1}.K^{-1}]}$ is the thermal conductivity, and $\sigma = (1 + j) / \mu$. In this equation, the first term of the denominator allows visualizing the two frequency regimes described in Sec.~\ref{subsec:pa_ir_s}.

Fig.~\ref{fig:results_plant} presents a numerical application of Eq.~\ref{eq:hu_homogeneous} to estimate the relative variation in $s$ modulus and phase as a function of the modulation frequency $f$, for a given variation in $\beta$. It is assumed that the sample is pure water, and that $\lambda = 9.65~\mathrm{\mu m}$.

\begin{figure}[htbp]
\centering
\includegraphics[width=1.00\linewidth]{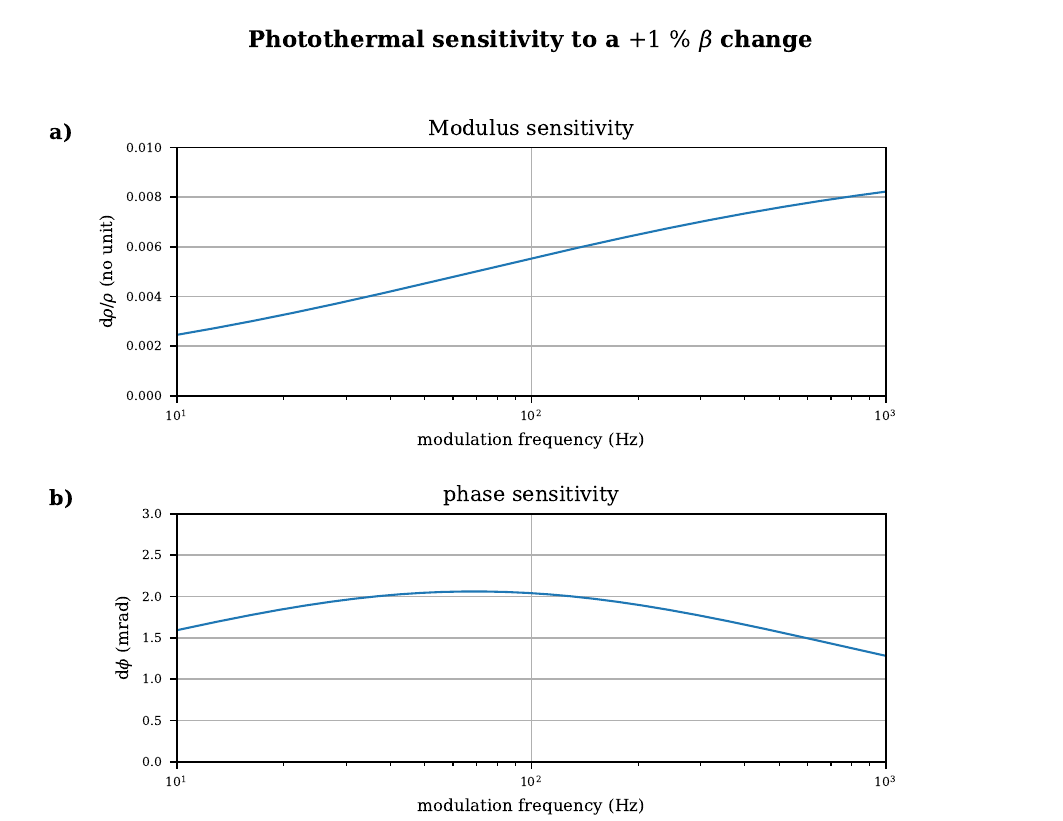}
\caption{Photothermal signal sensitivity to a $+1~\mathrm{\%}$ change in $\beta$. Subfigure \textit{a)} corresponds to the modulus, and \textit{b)} to the phase. The sample is homogeneous and made of water, the IR source wavelength is $\lambda = 9.65~\mathrm{\mu m}$.}
\label{fig:sensitivity_to_beta}
\end{figure}

\bibliographystyle{elsarticle-num-no-url}
\bibliography{manuscript_biblio}

\end{document}